\documentclass[aps,twocolumn,amssymb,prl,showpacs,10pt]{revtex4-1}

\usepackage[utf8]{inputenc}
\usepackage{tabularx}
\usepackage{multirow}
\usepackage{url}
\usepackage{amsmath}
\usepackage{amssymb} 
\usepackage{graphicx}
\usepackage{mathrsfs}
\usepackage{color}
\usepackage[normalem]{ulem}
\usepackage{float}
\usepackage{lmodern}
\usepackage{soul}
\usepackage[breaklinks,colorlinks,urlcolor=blue,citecolor=blue]{hyperref}
\usepackage{svg}
\usepackage{bibunits}

\usepackage{booktabs}
\usepackage{bm}

\newcommand{\unitspace}{\,}

\newcommand{\g}{\ensuremath{\unitspace \mathrm{g}}}

\newcommand{\Ten}[2]{\ensuremath{#1 \times 10^{#2}}}
\newcommand{\bfk}{\mathbf{k}}
\newcommand{\bfn}{\mathbf{n}}
\newcommand{\bfv}{\mathbf{v}}
\def\bfzeta{\bm{\zeta}}
\def\bfOmega{\bm{\Omega}}

\newcommand{\wasp}[0]{WASP-96\,b}
\newcommand{\hd}[0]{HD209458\,b}

\newcommand{\psection}[1]{\textit{\textbf{#1.---}}} 

\newcommand{\YK}{Y-K} 
\newcommand{\Th}{Th} 

\newcommand{\orcid}[1]{\href{https://orcid.org/#1}{\includegraphics[width=10pt]{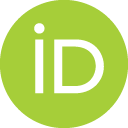}}}

\begin{document}
\begin{bibunit}

\title{Repeated Cyclogenesis on Hot-Exoplanet Atmospheres with Deep Heating}

\author{Jack W. Skinner$^{1,2\ast}$\orcid{0000-0002-5263-385X}, Joonas N\"attil\"a$^{3,4 \dagger}$\orcid{0000-0002-3226-4575}, James Y-K. Cho$^{2,3 \ddagger}$\orcid{0000-0002-4525-5651}}

\affiliation{$^1$ California Institute of Technology, 1200 East California Boulevard, Pasadena, California 91125, USA}
\affiliation{$^2$ Martin A. Fisher School of Physics, Brandeis University, 415 South Street, Waltham, Massachusetts 02453, USA}
\affiliation{$^3$ Center for Computational Astrophysics, Flatiron Institute, 162 Fifth Avenue, New York, New York 10010, USA}
\affiliation{$^4$ Physics Department and Columbia Astrophysics Laboratory, Columbia University, 538 West 120th Street, New York, New York 10027, USA}

\begin{abstract}
Most current models of hot-exoplanet atmospheres assume shallow heating, a strong day-night differential heating near the top of the atmosphere. Here we investigate the effects of energy deposition at differing depths in a model tidally locked gas-giant exoplanet. We perform high-resolution atmospheric flow simulations of hot-exoplanet atmospheres forced with idealized thermal heating representative of shallow and deep heating (i.e., stellar irradiation strongly deposited at $\sim 10^3$ \, Pa and $\sim 10^5$\,Pa pressure levels, respectively). Unlike with shallow heating, the flow with deep heating exhibits a new dynamic equilibrium state, characterized by repeated generation of giant cyclonic storms that move away westward once formed. The formation is accompanied by a burst of heightened turbulence, leading to the production of small-scale flow structures and large-scale mixing of temperature on a timescale of $\sim 3$ planetary rotations. Significantly, while effects that could be important (e.g., coupled radiative flux and convectively excited gravity waves) are not included, over a timescale of several hundred days the simulations robustly show that the emergent thermal flux depends strongly on the heating type and is distinguishable by current observations.

\end{abstract}

\maketitle

\psection{Introduction}
Current and upcoming space-based missions, such as the James Webb 
Space Telescope (JWST) \citep{Gardner06} and Ariel \citep{Tinetti21}, 
will enable next-generation observations of exoplanet atmospheres.
However, accurate dynamics modeling is required to interpret and guide 
these observations.
At present, most general circulation simulations of 1:1 spin--orbit synchronized 
hot-exoplanet atmospheres are performed with a strong 
day--night differential heating near the top of the modeled atmosphere \citep{Showetal09,RausMen10,ThraCho10,Dixonetal10,Hengetal11,LiuShow13,Maynetal14,Choetal15,Mend20, Showetal20, Choetal21,SkinCho22,Komaceketal22}.  
Here we refer to this type of heating as \emph{shallow heating}.
Importantly, \emph{deep heating}---i.e., a strong day--night differential heating 
at a much greater depth, in addition to near the top---has 
also been suggested in the past \cite{Showetal08,Dixonetal10,Gandi2020,Wellbanks2022,Changeat22}.  
Such variations in the energy deposition depth can be due to, e.g., changes in how the incoming stellar radiation is transferred, reflected, or absorbed by the planet's atmosphere.
In addition, a preliminary modeling of the recent JWST observation of the 
hot-exoplanet \wasp\ also suggests a deep heating for this 
planet~\cite{Note1}.
For exoplanets, the atmospheric response to such differences in 
heating is still not well understood \cite{Cho08,Choetal19}; hence, a 
robust study of the response is needed, for both theory and observation.
In this Letter we investigate the dynamical consequence of deep 
heating---specifically, stellar irradiation deposited at the pressure 
level of $\sim$$10^5$\,Pa, as well as at $\sim$$10^3$\,Pa---on 
hot-exoplanet atmospheres with a three-dimensional (3D), global 
hydrodynamics model.

In the case of the Earth, for which ample observations and modeling studies exist, it is well known that temperature profiles resulting from heating at different pressure levels (depths) lead to different large-scale dynamics.  This is due to, inter alia, thermal--mechanical forcing of lateral motion 
and entrainment/detrainment; see, e.g., \cite{Schumacheretal2004,Neggers2007,ZhangSong09,Zhao2014}, and references therein. 
Likewise, the equilibrium temperature distribution $T_e(\lambda,\phi,p,t)$ 
is a key factor that governs the large-scale dynamics on exoplanets; 
here $\lambda$, $\phi$, $p$ are the longitude, latitude, and pressure, 
respectively; and, $T_e$ is generally dependent on time~$t$, as indicated. 
Under 1:1 spin--orbit synchronization, in the absence of atmospheric motion, 
a permanent dayside and nightside is expected on the planet---especially 
at low $p$-levels.  
However, dynamics can subvert this expectation \cite{Choetal03,Choetal08,ThraCho10,WatCho10,Choetal15,Choetal21,SkinCho22,Komaceketal22}.
For example, with deep heating, heavier masses of atmospheric fluid 
can be energized because of stratification \cite{Choetal03} and stronger 
upwardly-propagating internal gravity waves are excited \cite{WatCho10}.
Both mechanisms can smooth, or even reverse, the dayside--nightside 
temperature gradient via advection and wave--mean flow interaction.
In this way, dynamics is crucial in configuring the temperature 
field, in both space and time---and ultimately what is observed.

\psection{Numerical Simulation}
The global dynamics, and its associated temperature distribution, of 
hot-exoplanet atmospheres are simulated by solving the primitive 
equations~(PEs) in the vorticity--divergence--potential temperature 
form \cite{Scott2004BOB:Guide,Polietal14}
 \begin{eqnarray}
    \frac{\partial{\zeta}}{\partial t}\ & = & 
     \bfk \cdot (\nabla \times \bfn) \nonumber \\ 
    \frac{\partial{\delta}}{\partial t}\ & = & 
     \nabla \cdot \bfn\ -\ 
     \nabla^2\!\left(\frac{1}{2}\,\bfv^2 + \Phi\right)\ \nonumber \\ 
     \frac{\partial{\Theta}}{\partial t} & = &   
     -\nabla\!\cdot\!\Big(\Theta\,\bfv\Big)\ -\ 
     \frac{\partial\ }{\partial p} \Big(\omega\,\Theta\Big)\ +\ 
     \frac{\dot{q}_{\rm net}}{\Pi}\, ,  
 \end{eqnarray} 
where decreasing $p$ is the increasing vertical coordinate, $\zeta \equiv \bfk\cdot\nabla\times\bfv$ and 
$\delta \equiv \nabla\cdot\bfv$ are the vorticity and divergence of the horizontal
velocity field $\bfv(\lambda,\phi,p,t)$, respectively, and 
$\Theta \equiv (c_p/\Pi)\, T$ is the potential temperature---related 
to the entropy $s$ by ${\rm d}s = c_p\, {\rm d}(\ln\Theta)$.  
Here $\bfk$ is the local vertical direction (magnitude of unity); 
$\nabla$ is the horizontal gradient operator on constant $p$-surfaces; 
$c_p$ is the specific heat capacity at constant $p$; 
$\Pi \equiv c_p\,(p/p_0)^{{\cal R}/c_p}$ is the Exner function, 
where $p_0$ is a constant reference pressure 
($\equiv 10^6$\,Pa in this work) and 
${\cal R}$ is the mean specific gas constant; 
and, $T(\lambda,\phi,p,t)$ is the ordinary temperature field. 
Also, $\bfn\ \equiv\ -(\zeta + f)\,\bfk\!\times\!\bfv\, -\, 
\delta\,\bfv\, -\, \partial(\omega\bfv)/\partial p$, where $f$ is the 
Coriolis parameter and $\omega \equiv {\rm d} p / {\rm d} t$ is the 
vertical velocity; 
$\Phi$ is the geopotential; and, 
$\dot{q}_{\rm net}(T_e)$ is the net heating rate field, the key element 
in this work.  

The Eq.~(1) set is closed with the following relationships: 
$\partial \Phi / \partial \Pi =  -\Theta$, 
$\partial \omega / \partial p = - \delta$,
and $p = \rho RT$, 
where $\rho$ is the density field.  
The boundary conditions at the top and bottom of the domain are 
free-slip---i.e., $\omega = 0$.  
As for the lateral direction, the fields are periodic in $\lambda$, 
and $\bfv\cos\phi$ mapping is used to represent $\bfv$ in terms of 
the scalar Legendre expansions.
The latter is convenient for solving the equations 
\cite{Scott2004BOB:Guide} and forces $\bfv\cos\phi$ to be null 
at the poles.

\begin{figure}
\centering
\includegraphics[trim={0.0cm 0.0cm 0.0cm 0cm}, clip=true, width=0.45\textwidth]{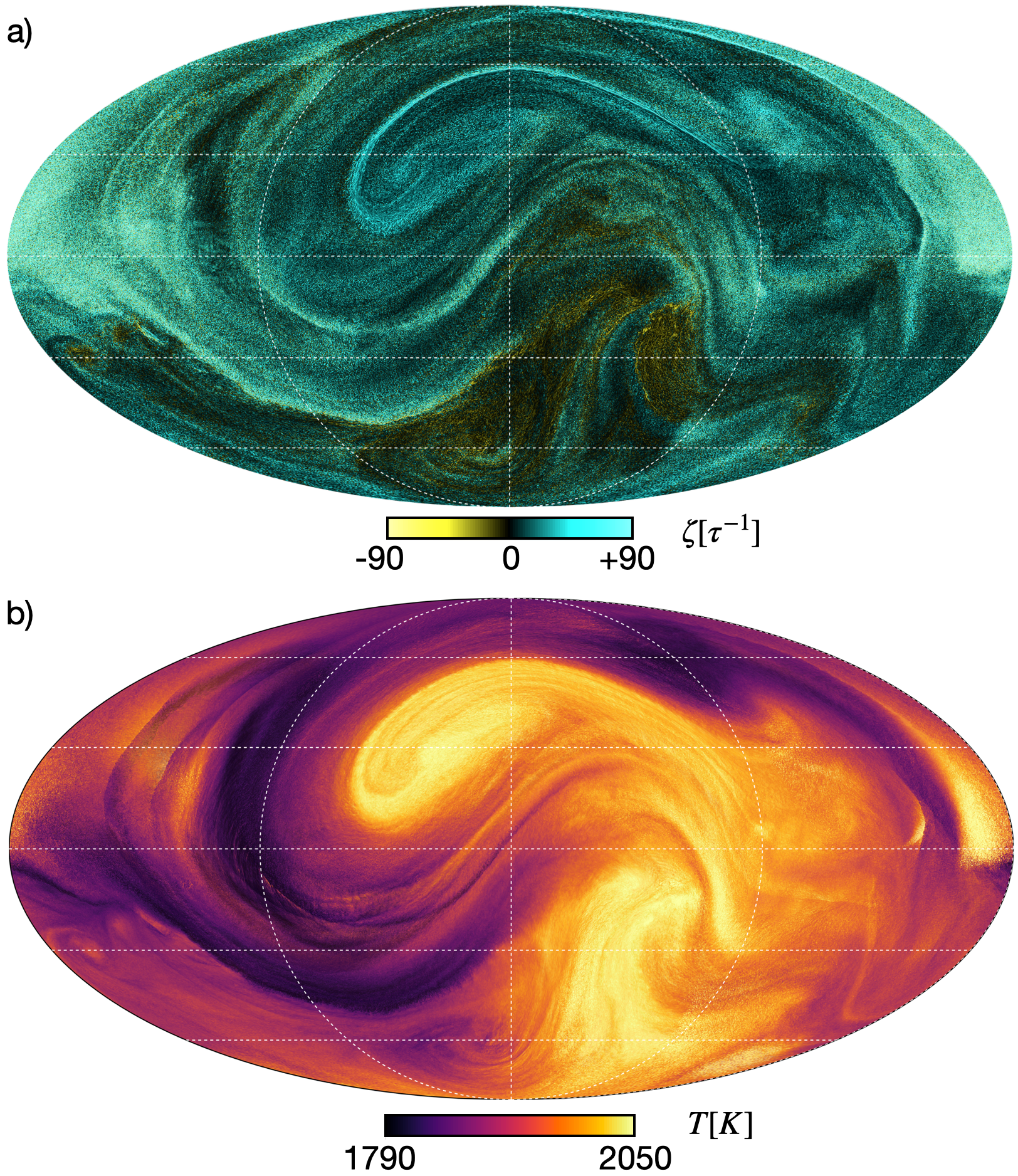}\\
\caption{
Vorticity $\zeta$ and temperature $T$ fields (a and b, respectively) 
with deep heating at time $t = 155$ and at the $10^5$\,Pa pressure level.
The generic response to deep heating is large-scale vortices rolling up, breaking off, 
and then moving westward away from the point of formation 
at mid-latitudes.   The point may be slightly shifted in longitude at 
other times or in simulations with slightly different numerical 
parameters; a movie is available~\cite{youtube_link}.
\vspace{-0.5cm}
}
\label{fig:fig1}
\end{figure}

The full set of PEs is solved with a well-tested pseudospectral code 
employing hyperviscosity \cite{SkinCho21,Polietal14,Choetal15}.  
The results are based on numerically converged, as well as 
consistent, simulations \cite{Strikwerda2004} at very high 
resolution---i.e., a horizontal resolution of up to T1365 
(corresponding to a truncation of 1365 total modes and 1365 azimuthal 
modes in the Legendre expansion) and  with a vertical resolution of up 
to L200  (corresponding to 200~layers in $p$-space) at T682 resolution.
A snapshot from a sample T1365L20 solution is presented in 
Fig.~\ref{fig:fig1}.
Note the intricate folding and layering of fine-scale fronts around 
the large organized $\zeta$ and $T$ areas, which are constantly in 
motion; note also the close relationship between the flow and 
temperature.
The fine-scale structures, which cannot be captured without high 
resolution and low dissipation, are paramount for accurate numerical 
solutions and observation interpretations \cite{SkinCho21}.
Other important numerical parameters, along with their values, and 
convergence tests of the simulations are provided in the Appendixes; see also  Refs.~\cite{Choetal15},  \cite{Polietal14}, 
and \cite{SkinCho21} for broader comparisons.

In this Letter we discuss two exoplanets, {\hd} and {\wasp}.  
Although the precise shape of temperature profiles for any exoplanet 
is currently unknown, the two serve as fiducial examples of exoplanets 
with atmospheres characterized by shallow heating and deep heating, respectively \citep[][]{Note1, Showetal09}.
Apart from the profiles, they possess physical parameter values that 
are effectively the same, from an atmospheric dynamics standpoint.
Hence, they permit the effects of heating type on the dynamics to be 
well-delineated.  
The relevant physical (orbital and atmospheric) parameters, and their 
values, are also given in the Appendixes.
The characteristic advection time, $\tau_a \equiv R_p / U$, is similar 
for both planets; here $R_p$ is the planetary radius and $U$ ($\sim\! 10^3$\,m\,s$^{-1}$) is the characteristic speed of a structure of length 
scale $R_p$.
Additionally, for large portions of the atmosphere on both planets 
$\tau_a \ll \tau$, where $\tau$ is the planetary rotation period. 
From hereon time is given in terms of $\tau$.

\begin{figure}
    \centering
    \includegraphics[width = 0.45\textwidth]{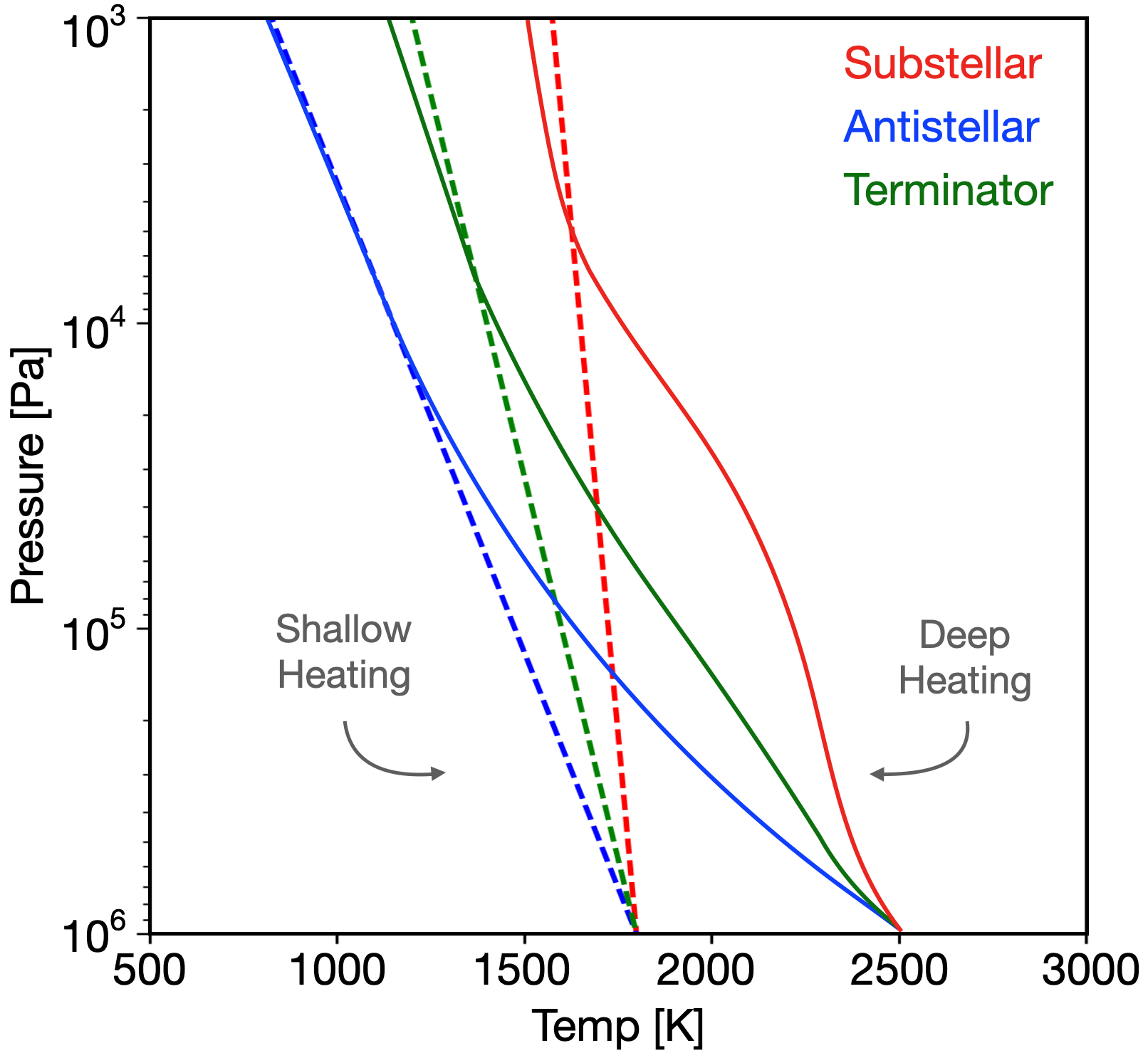}
    \caption{
    Temperature profiles $T(p)$ of shallow heating (dashed lines) and deep heating (solid lines).  Colors refer to different points in longitude.  The dayside--nightside difference is larger and the thermal relaxation time is smaller in the deep region for deep heating.
    \vspace{-0.5cm}
    }
    \label{fig:fig2}
\end{figure}

The salient difference between the two planets is the strength 
(amplitude and timescale) of the thermal forcing in the deep 
region of their atmospheres---i.e., at $p$-levels between 
$\sim$$\Ten{5}{4}$~Pa and $\sim$$\Ten{5}{5}$~Pa.
Fig.~\ref{fig:fig2} shows the model $T(p)$ profiles for the planets, these 
are derived from one-dimensional (1D) radiative transfer calculations 
that incorporate Spitzer and Hubble observations \cite{Changeat22,Changeatetal22, Showetal09}; 
all profiles represent stable stratification throughout the modeled atmosphere.  
In this work, both atmospheres are driven purely by a thermal relaxation 
to a prescribed temperature distribution ($\propto \cos\lambda\cos\phi$ 
on the dayside and uniform on the nightside) and on a timescale $\tau_r$ ($\propto\, p\, /\, T_e^4$)~\cite{Salby1996}.
Although highly idealized, the forcing is reasonable---particularly for 
the deep region~\cite{Choetal08}.
Note that, in the deep region $\tau_r / \tau_a \gtrsim 1$ for \hd\ \cite{Choetal03,LiuShow13,Choetal15} 
whereas $\tau_r / \tau_a \lesssim 1$ 
for \wasp; in the upper region (i.e., $p \lesssim 10^4$\,Pa), 
$\tau_r / \tau_a \ll 1$ for both planets.  
The larger dayside--nightside $T$ difference and shorter $\tau_r$ in 
the deep region of \wasp\ lead to a significant difference in the flow 
and temperature distributions, as shown below.  

\psection{Modon Pair with Shallow Heating}
On 1:1 spin--orbit synchronized hot-exoplanets, the intense irradiation 
from their host stars creates flow structures on a wide range of scales 
and amplitudes \cite{Choetal21,SkinCho21}.
These include planetary-scale vortex-couplets, small-scale vortex singlets, 
large-amplitude Rossby waves, and meso-scale gravity waves \cite{Choetal03,WatCho10,Choetal15,Choetal21}.  
They are generic to planets with strong dayside--nightside differential 
heating mostly in the upper region \cite{ThraCho10,SkinCho21}.  

\begin{figure}
    \includegraphics[width = 0.45\textwidth]{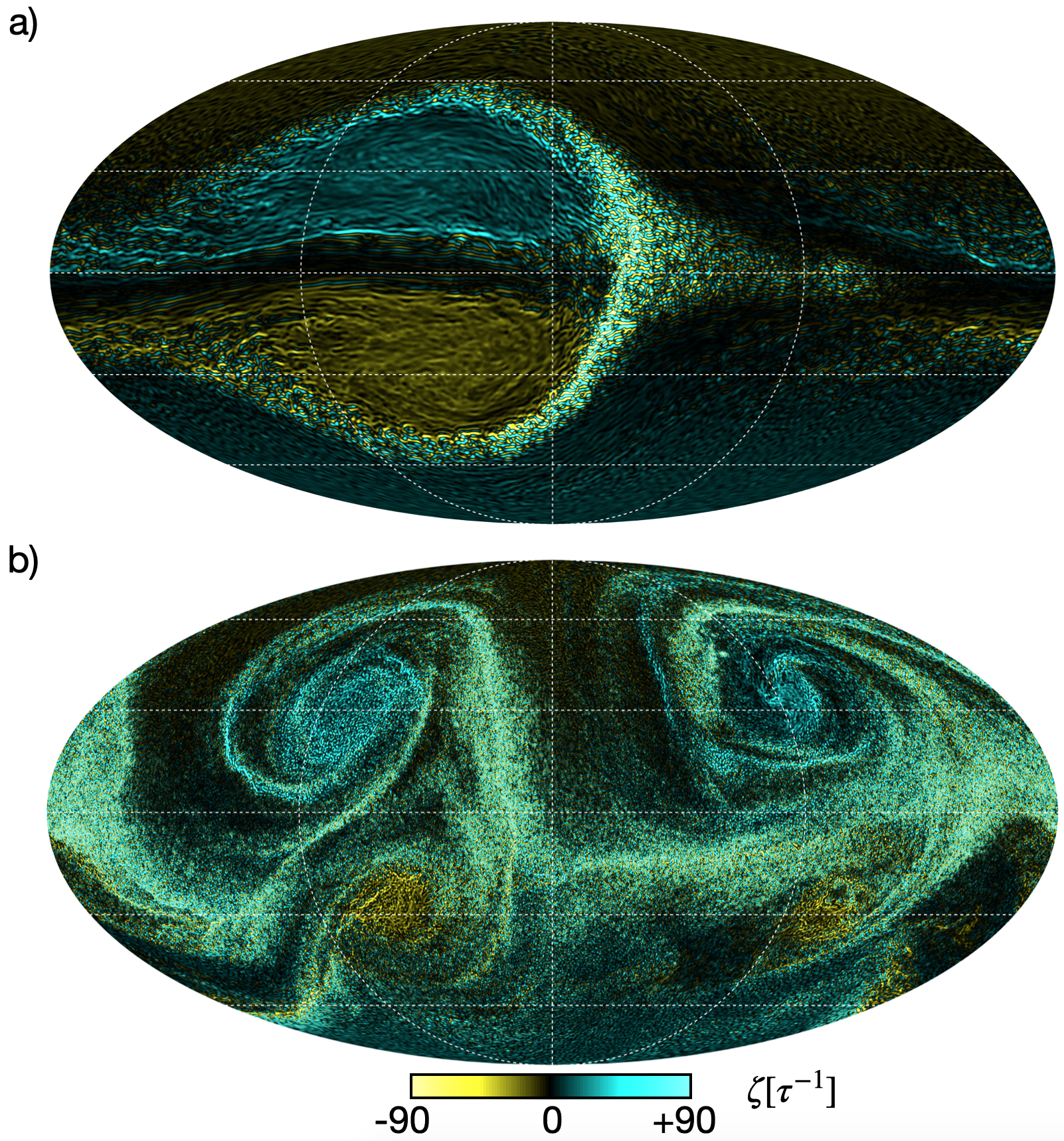}
    \caption{
    $\zeta$ fields at the $10^5$\,Pa level with shallow and deep heating (a and b, respectively), showing the flows dominated by different organized structures (after $t \approx 10$).
    In a) the flow is dominated by a modon and a much weaker antimodon, whereas in b) the flow is dominated by a quartet of cyclones in a von K\'arm\'an vortex street-like formation.  In the latter, the cyclone quartet exhibits a quasi-periodic life cycle (of period $\sim$10) that begins with the flow similar to that seen in  Fig.~\ref{fig:fig1}a.}
    \vspace{-0.5cm}
    \label{fig:fig3}
\end{figure}

Under shallow heating, giant modons---vortex couplets with their 
constituent vortices having opposite signs of 
vorticity \cite{HogStom85}---are the preeminent structures that emerge 
in the flow; see Fig.~\ref{fig:fig3}a.  
Generally, there are two of them---a modon and an antimodon (composed of 
two cyclones and two anticyclones that initially span across the equator, respectively): 
a cyclone is a vortex with $\bfzeta\cdot\bfOmega > 0$ and 
an anticyclone is a vortex with $\bfzeta\cdot\bfOmega < 0$, where 
$\bfzeta = \zeta\bfk$ and $\bfOmega$ is the planetary rotation vector 
that orients the north pole.
These long-lived, non-stationary structures are responsible for 
transporting and mixing hot {\it and} cold patches of the atmosphere, 
in their cores as well as in their interstitial areas.  
In this way, they induce distinct time-varying thermal signatures that 
could be observable \cite{Choetal03,Choetal21,SkinCho22}.
Here small-scale flow structures are important because they perturb 
the giant modons from stationary or smoothly-translating states, seen 
in lower resolution or more dissipative simulations \cite{SkinCho21}.  
In general, modons move chaotically, due to the continual nonlinear 
interactions with the small-scale structures \cite{SkinCho22}.  
Over long duration, the flow evolution exhibits several equilibrium 
states, with occasional sudden transitions between 
them \cite{Choetal21}.  
These states are: \textit{i}) steadily translating, \textit{ii}) 
oscillatory, and \textit{iii}) vortex-street-like.  

In state~\textit{i}), a strong modon and a much weaker antimodon form in the deep region near the substellar and antistellar points, respectively.  Both couplets eventually drift westward after an initial eastward translation by the modon.  
Subsequently, one or both may spread apart or entirely break up repeatedly near the eastern terminator.  
In state~\textit{ii}), the two cyclones of the modon spin out alternatively in the northern and southern hemispheres, both moving off to the west after spinning out.  
This leads to a noticeable symmetry breaking between the two hemispheres.  
Importantly, this state is transient, lasting a much shorter duration than the other states. 
In state~\textit{iii}), cyclones straddle a large-amplitude equatorial Rossby wave and steadily translate westward with the wave in a von K\'arm\'an vortex street-like chain \cite{KarmanRubach12}---without breaking up.   This state is generally seen in the very deep regions ($p\, \gtrsim\, 10^6$\,Pa), post a very long build-up period (typically $t \sim 200$)~\cite{Choetal21}.

\psection{Cyclone Quartet with Deep Heating}
Fig.~\ref{fig:fig3}b shows the $\zeta$ field for \wasp\ at $t = 157$ (at 
the same $p$-level of the fields in Fig.~\ref{fig:fig1}).  
Fig.~\ref{fig:fig3}b should be compared with Fig.~\ref{fig:fig1}a, as the 
two give an idea of the dynamic and quasi-cyclic nature of the flow with 
deep heating:  they illustrate roughly the half points of a cycle. 
That is, the field shown in Fig.~\ref{fig:fig3}b is essentially the same 
as that at $t \approx 145$ (as well as at $t \approx 15$), and the field 
shown in Fig.~\ref{fig:fig1}a is essentially same as that at 
$t \approx 165$ (albeit with $\sim$\,$45^\circ$ westward shift in the 
latter).  

In general, the flow contains a much greater number of small-scale 
vortices at the shown $p$-level than that with shallow heating 
(cf. Fig.~\ref{fig:fig3}a and Fig.~\ref{fig:fig3}b).  
More importantly, unlike in simulations with shallow heating, large 
uncoupled cyclones---usually four of them---repeatedly form and 
individually spin off westward at mid-latitudes.
This occurs after an initial brief period of a giant cyclonic modon 
formation, subsequent rapid eastward translation, and total breakup 
on the dayside ($t \lesssim 10$).  
After the breakup, large cyclones form via Rossby wave breaking 
in both the northern and southern hemispheres: the wave breaks in 
the counter-clockwise and clockwise directions in the northern and 
southern hemispheres, respectively, rolling up into cyclones.  
The growth time for the formation is very short ($\sim$\,1); 
but, once formed, the cyclones last for a period of 
$\sim$\,3 to $\sim$\,15, before succumbing to large-scale 
external straining.  
The signature of the cyclone quartet life cycle can be seen in 
disk-integrated thermal flux, as discussed below.  
The copious small-scale vortices (storms), continuously regenerated 
in bursts and generally populated along the sharp fronts surrounding 
the cyclones, also remain coherent for a long time ($\,\gtrsim\! 10$).  

\begin{figure}
\centering
\includegraphics[width = 0.48\textwidth]{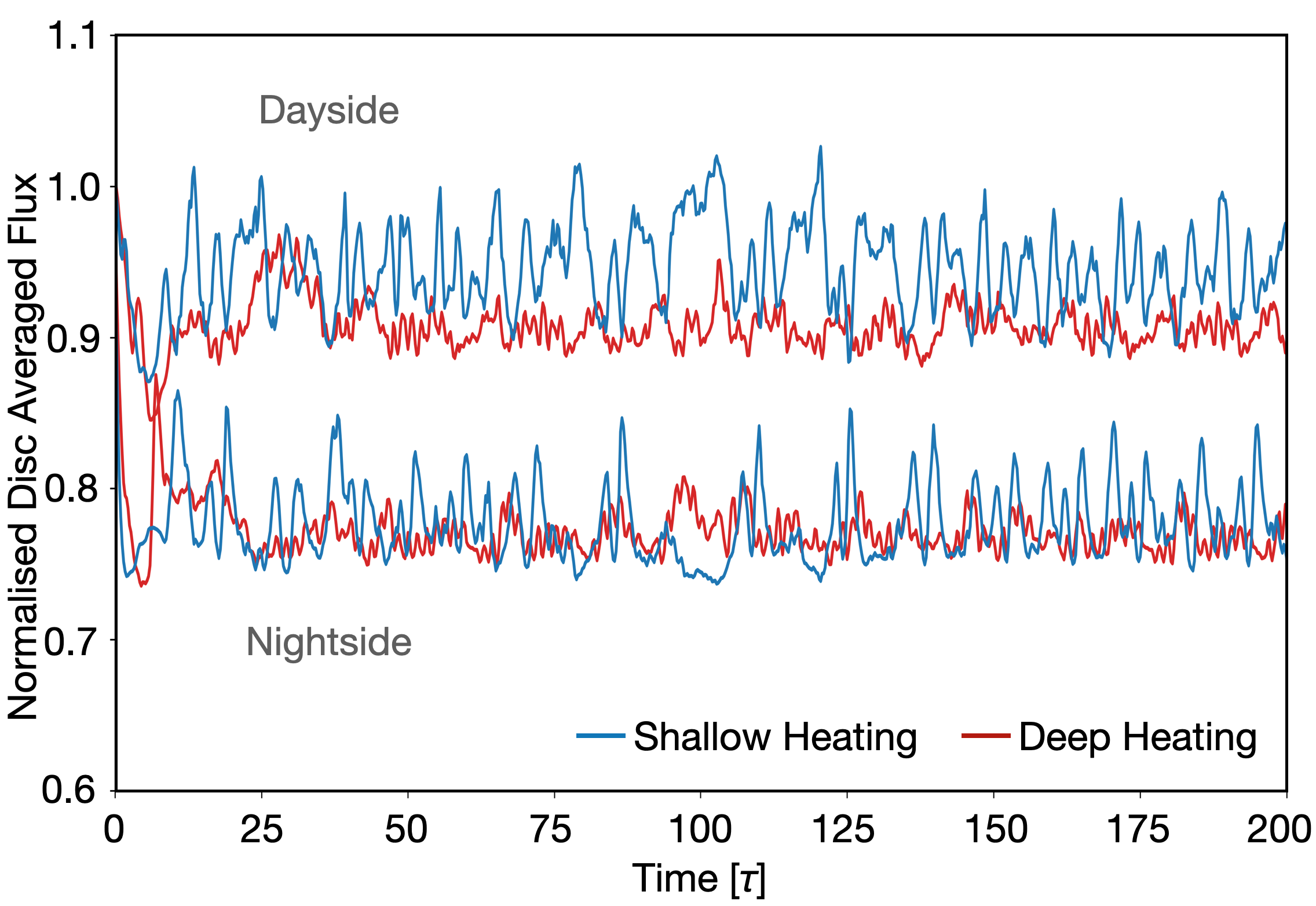}
\caption{
    Time-series of disk-averaged thermal flux ($\propto \sigma T^4$, where $\sigma$ is 
    the Stefan--Boltzmann constant) at the $p = 10^5$\,Pa level, with disks centered 
    on the dayside (top) and nightside (bottom), for atmospheres with shallow heating 
    (blue) and deep heating (red).  Each series is normalized by the initial terminator 
    value at the 
    $p = 10^5$\,Pa level. 
    \vspace{-0.5cm}
}
\label{fig:fig4}
\end{figure}

The recurring cyclogenesis and bursts of turbulence vigorously mix 
the atmosphere and homogenize the planet's $T$ field over short 
timescales ($\sim$3); see Fig.~\ref{fig:fig4}---in particular, the 
much smaller peak amplitudes compared to those for the shallow heating 
case.
The geneses and bursts cause the hot and cold patches of the atmosphere 
to be highly variable in space and time.
While the flow of state~{\it ii}) with shallow heating is moderately 
similar, that flow is much less dynamic (with greatly reduced $U$ and turbulence) {\it and} 
occurs at much greater depths (i.e., $p \gtrsim 10^6$\,Pa) \cite{Choetal21}.
This is due to the markedly different forcing amplitude and relaxation timescale
in the deep region between the two types of heating, as they 
power different mechanisms for vortex formation  (e.g., barotropic 
instability with shallow heating and baroclinic 
instability \cite{PoliCho12} with deep heating).

Most importantly, this ``cyclogenesis plus enhanced turbulence'' state 
is the {\it sole} equilibrium state exhibited with deep heating---i.e., multiple 
equilibrium states are not exhibited, as with shallow heating.
Additionally, with shallow heating, a weak antimodon initially forms in 
the region $p \lesssim \Ten{2.5}{4}$\,Pa (see Fig.~4 of 
Ref.~\cite{SkinCho22}); a much weaker, antisymmetric version of this 
structure is present with shallow heating on the nightside and together constitute
a pair of heton quartets \cite{Kizner06}.
After its formation, the cyclonic modon slowly becomes more columnar (barotropic) over time. 
However, with deep heating, the four cyclones are strongly barotropic (extending 
essentially over the entire region above $p \approx 10^5$\,Pa) 
from the time they first begin to emerge at $t \approx 10$.   
After $t \approx 17$, the cyclones are repeatedly generated throughout 
the simulation duration  ($t = 300$ here and longer at lower resolutions).
We emphasize that baroclinic instability mainly powers the cyclone 
generation; overall, the flow is strongly barotropic.

Given the marked difference in behavior, the type of heating on 
hot-exoplanets could be constrained by current observations.
Consider, for example, the time series of disk-averaged thermal flux, 
corrected with a cosine projection factor for the geometry.
Four series are shown in Fig.~\ref{fig:fig4}, with the disks centered on 
the antistellar and substellar points (labeled nightside and dayside, respectively).
The nightside series corresponds to the flux at the peak primary eclipse, 
and the dayside series corresponds to the flux just before ingress or 
after egress of the secondary eclipse. 
The series are from the simulations presented in Fig.~\ref{fig:fig3} 
with the shallow heating case in blue and the deep heating case in 
red. 

In both cases, the motion of large storms around the planet leads to 
flux peaks that are roughly $180^\circ$ out of phase at the dayside 
and the nightside.
However, as already noted, the amplitude of the peaks is significantly 
reduced with deep heating compared to those with shallow heating.  
This difference arises because the average over the disk is modulated 
by the homogeneity of the $T$ field, caused by the mixing.  
With deep heating, hot and cold patches periodically peel away 
westward from their points of emergence on the dayside and nightside, 
respectively---often past the terminators.
This results in a flux series with a lower amplitude and higher 
frequency peaks.   
At $p\!\sim\! 10^5$\,Pa, the hottest patch of the atmosphere is 
generally situated near the substellar point on the dayside, and even 
slightly shifted eastward at times; storms that spin off do sequester 
and transport hot temperatures towards the western terminator at 
mid--high latitudes but, tempered by the viewing geometry, do not 
contribute significantly to the disk-averaged flux centered on the 
substellar 
point.
Superficially, this behavior appears similar to that reported in 
past simulations employing shallow heating and low resolution; 
however, the behavior here is much more dynamic and is generated by 
a completely different mechanism.
With shallow heating, hot patches advected by a modon 
moving quasi-periodically around the planet over a longer period lead 
to high-amplitude and lower frequency peaks in the flux.
As a consequence, the shallow heating case would appear to have a 
more westward shifted ``hot spot'' than the deep heating case.
Note that the difference in the peaks between the two types of 
heating ($\sim$100\,K, or $\sim\! 4$\% of the prescribed  
temperature at the substellar point) is greater than the expected 
accuracy of JWST measurements, which is $\approx\! 30$\,K 
\cite{JWST_com, Bean_2018}.

\psection{Discussion} 
In this letter we investigate the effects of different depths of heating on 
the atmospheric dynamics of exoplanets. Two generic types of heating are 
heuristically designated 
as ``shallow heating'' and ``deep heating".
While we remain agnostic on the exact origin of the energy deposition 
variations, we note that such changes are naturally expected, e.g., due to 
changes in how the host star radiation is transferred and absorbed in the
planet's outer layers.
Such changes are further amplified by the unknown vertical distribution 
of clouds and aerosols in the atmosphere.
Interestingly, 
both types of heating are consistent with temperature profiles obtained by 1D radiative transfer calculations 
utilizing Spitzer, Hubble, and JWST observations \cite{Gandi2020, Wellbanks2022, Changeatetal22, Changeatetal23}.

Here, the focus is laid on the influence of the 
deep
heating---particularly on the large-scale atmospheric dynamics and its associated temperature redistribution in the modeled region, $p\in[10^3,10^6]$\,Pa.
In the past, most simulations have employed shallow heating only.
Some physical mechanisms that may influence the dynamics and temperature 
(e.g., the radiative flux feedback from temperature redistributions and saturation of upwardly propagating gravity waves excited by convection 
below) have not been included; but, they are not expected to qualitatively change the main result reported here because of the duration and vertical 
range of the simulations (\cite{Choetal19}, and references therein).

The main result here is that, with deep heating, a new persistent 
flow state arises in which giant cyclones repeatedly form near the 
substellar point and move away westward.
The formation and subsequent motion redistribute the temperature more homogeneously over the planet, compared with the shallow heating case.
Importantly, the cyclones---generally a quartet of them---are 
accompanied by bursts of heightened turbulence activity.
This enhances the mixing of large hot and cold areas of the atmosphere. 
Moreover, unlike with shallow heating, this is the sole equilibrium state exhibited with deep heating; with shallow heating, a multitude of states 
is exhibited.

The new, exclusive state exhibited in the flows driven by deep 
heating is characterized by disk-averaged temperature fluxes which are 
overall less variable, particularly near $p = 10^5$\,Pa.
These fluxes are lower in amplitude, compared with those generated by 
flows driven by shallow heating. 
More specifically, the typical difference in the flux is $\sim$$10\%$.
This is produced by a disk-averaged temperature difference of 
${\cal O}(10^2\,{\rm K})$ between the atmospheres with the two types of 
heating.
In addition to the amplitude difference, the thermal flux also exhibits 
markedly different temporal behavior for the two types of heating;
for example, the fluxes for deep heating contain higher frequency 
perturbations, compared with those for shallow heating.
Such differences are distinguishable by next-generation telescopes 
such as the JWST \cite{Gardner06} and Ariel \cite{Tinetti21}.
Therefore, the two types of heating can be clearly delineated by 
current observations.

\vspace{1em}
\psection{Acknowledgements}
The authors thank Quentin Changeat for helpful discussions. 
J.W.S thanks the Jet Propulsion Laboratory, California Institute of Technology, where some of this work was completed under a contract with the National Aeronautics and Space Administration (80NM0018D0004).
J.N is supported by a joint Columbia University and Flatiron Institute Research Fellowship.
The Flatiron Institute is acknowledged for providing computational 
resources and support.
This work resulted from an initial discussion at the Exoplanet Symposium 
2022, organized at the Flatiron Institute. 

\vspace*{0.5cm}
\hspace*{0.1cm} $^\ast$~jskinner@caltech.edu\\
\hspace*{0.45cm} $^\dagger$~jnattila@flatironinstitute.org\\
\hspace*{0.45cm} $^\ddagger$~jamescho@brandeis.edu 

\vspace{-1.4cm}

\appendix
\clearpage
\newpage

\section{A. Numerical Model}
In this work, we employ the pseudospectral code used in Refs.~\cite{Polietal14, SkinCho21, Choetal21} to simulate the atmospheric dynamics of tidally synchronized hot-exoplanets, \hd\ and {\wasp}.
In this section, we provide additional technical information pertaining to the simulations.
More details are given in Refs.~\citep{Polietal14, SkinCho21}.
The code solves the hydrostatic primitive equations in vorticity--divergence--potential temperature form \citep{Scott2004BOB:Guide} using pressure $p$ as the vertical coordinate.
The equations governing the dynamics are integrated on a sphere providing natural boundary conditions in the horizontal ($\lambda, \phi$) directions (where $\lambda$ is the longitude and $\phi$ is the latitude).
In the vertical direction, we employ the free-slip condition  (i.e., ${\rm D} p / {\rm D} t = 0$, where $\mathrm{D}/\mathrm{Dt}$ is the material derivative) at the top and bottom $p$-surfaces.

The initial condition and thermal forcing employed in the \hd\ simulations are the same as in previous works \cite{Choetal15, SkinCho21, SkinCho22}.
The flow is forced with a Newtonian relaxation scheme \cite{Salby1996}, which relaxes the terminator temperature distribution (given by the average of the substellar and antistellar equilibrium temperatures) to the specified equilibrium temperature distribution.  The substellar, antistellar, and terminator profiles, all as functions of $p$, are as shown in Fig.~2 of the main text.
These profiles are derived from radiative transfer calculations of {\wasp} \footnote{Changeat, Q. (private comm.)} and {\hd} \cite{Showetal09}. 
The timescale for radiative cooling $\tau_{\rm rad}(p)$ is computed from for each of the terminator profiles following \cite{Choetal08};  $\tau_{\rm rad}(p)$ ranges from $\mathcal{O}(10^6~\!{\rm s})$ at $p = 10^6$ Pa to $\mathcal{O}(10^4~\!{\rm s})$ at $p = 10^3$ Pa. 
We have verified that the profiles we use are stably stratified by computing the Brunt-V\"ais\"al\"a frequency $\mathcal{N}$ and confirmed that $\mathcal{N}^2 > 0$ throughout the simulation domain.

\section{B. Physical and Numerical Parameters}

The physical parameters for the two planets discussed in the main text are summarized in Table~\ref{tab:params}.
Additional parameters for these planets can also be found in Refs.~\cite{Hellier2014} and \cite{ThraCho10}, respectively. 
Numerical parameters for the simulations of {\wasp} and {\hd} are summarized in Table~\ref{tab:num_params}. 

In this work, we denote the simulation resolution (e.g., T341L200) by the ``T number'' for the horizontal resolution, which is the truncation wavenumber in the spherical harmonic expansion, and the ``L number'' for the vertical resolution, which is the number of (linearly spaced) layers in the $p$-space.
Regarding the latter,
$p_t$ (always set to be 0) and $p_b$ are the top and bottom boundaries of the computational domain, respectively.  The prognostic variables (vorticity, divergence, and potential temperature) are defined on ``$1/2$-layers'', referred to as active layers.  Hence, $\tilde{p}_t$ and $\tilde{p}_b$, which are the topmost and bottommost active levels, are independently varied while $p_t$ and $p_b$ enforce the free-slip boundary conditions.

The generalized diffusion operator $\nabla^{2\mathfrak{p}}$ depends on the order $\mathfrak{p}$; here
$\mathfrak{p} = 1$ is the usual Laplacian operator and $\mathfrak{p}=8$ , for example, is the 16$^{\rm th}$-order hyper-viscosity operator \cite{ChoPol96a}.
The operator is accompanied by the related viscosity coefficient $\nu_{2\mathfrak{p}}$ multiplying the operator.
Lastly, we use a Robert--Asselin filter \cite{Asselin72} with coefficient $\epsilon$ \cite{ThraCho11}. 

The simulations discussed in the main text are of resolution
T1365L20 with 
$\tilde{p}_t = 5 \times 10^3 \,\mathrm{Pa}$,
$\tilde{p}_b = 1 \times 10^6 \,\mathrm{Pa}$,
$\Delta t = 4.5 \,\mathrm{s}$, 
$\mathfrak{p} = 8$, 
$\nu_{16} = 3.5 \!\times\! 10^{-53}\,R_p^{16}\,\tau^{-1}$, and  $\epsilon=0.02$.
The duration for the simulations is 300\,$\tau$, long enough for the system to reach quasi-equilibrium and exhibit more than 10 cycles of cyclone generation thereafter.  We have verified that the flow evolution in the simulations is qualitatively the same as those in longer duration simulations, at lower horizontal resolutions---i.e., 500\,$\tau$ and 1000\,$\tau$ at T682L50 and T341L200, respectively.

\begin{table}[t]
\raggedright 
\setlength{\tabcolsep}{0.30em}
\def\arraystretch{1.5}
\begin{tabular}{llll}
\hline
Parameter                  & HD209458b                          & WASP-96b                 & Units \\ 
\hline\hline
Rotation period  $\tau$   & 84.0                              & 81.6                    & hours   \\
Rotation rate $\Omega$    & $2.08\!\times\! 10^{-5}$                & $2.14\!\times\! 10^{-5}$    & s$^{-1}$    \\
Radius $R_p$              & $1.00 \!\times\! 10^8$                 & $8.58 \!\times\! 10^7$       & m \\
Mass $M_p$                & $1.31 \!\times\! 10 ^{27}$             & $9.11 \!\times\! 10^{26}$    & kg \\
Surface gravity $g$       & 10.0                                   & 8.27                         & ms$^{-2}$\\ 
Specific gas const. $\mathcal{R}$  & $3.50\!\times\!10^3$          & $3.50\!\times\! 10^3$  & ${\rm J\,kg}^{-1}{\rm K}^{-1}$\\ 
Specific heat const. $c_p$  & $1.23\!\times\! 10^4$                & $1.23\!\times\! 10^4$        & ${\rm J\,kg}^{-1}{\rm K}^{-1}$ \\
$T_{\rm SS} \, (T_{\rm AS})$ at $10^3$\, Pa & 1570 (820) & 1550 (820) & K \\
$T_{\rm SS} \, (T_{\rm AS})$ at $10^6$\, Pa & 1800 (1800) & 2500 (2500) & K \\
\hline
\end{tabular}
\caption{Physical parameters of the exoplanets, {\hd} and \wasp, used for the simulations in the main text.}
\label{tab:params}
\end{table}

\section{C. Convergence Tests}

The simulations in this work have been extensively tested and verified for numerical consistency and convergence.
By varying all the parameters listed in Table~\ref{tab:num_params}, the results have been verified to remain qualitatively unchanged.
Specifically, we have performed effectively identical (i.e., numerically equatable) simulations with increasing horizontal resolution, to confirm that the features of the flow remain qualitatively the same.
We note that the solutions are converged only at T341L10 resolution \cite{SkinCho21}.  
We have also verified that the primary conclusions do not change when $\tilde{p}_t$ and $\tilde{p}_b$ are independently varied while keeping L per Pa constant; see Table~\ref{tab:num_params}.
In particular, we have verified that varying $\tilde{p}_b$ away from $10^6$\,Pa, the level to which the thermal forcing extends, does not qualitatively change the salient features of the flow and temperature fields discussed in the main text (e.g., cyclone generation, thermal waves, and sharp fronts).
Varying $\tilde{p}_t$ away from $5\times 10^3$\,Pa, roughly the level to which simulations typically extend \cite{Polietal14,LiuShow13,Choetal15,SkinCho21}, also does not qualitatively change the salient features of the flow and temperature fields---particularly in the deep regions ($5\times 10^4\,{\rm Pa}\, \lesssim\, p\, \lesssim 10^6\,{\rm Pa}$), discussed in this work.

\begin{table}[t]
\raggedright 
\setlength{\tabcolsep}{0.2em}
\def\arraystretch{1.5}
\begin{tabular}{ll}
\hline
Parameter   & Value     \\ 
\hline\hline 
Horizontal resolution ${\rm T}$  & \{341, 682, 1365\}     \\
Vertical resolution ${\rm L}$    & \{3, 10, 20, 50, 100, 200\}      \\
Top active $p$-level $\tilde{p}_t$\ [$10^3\,{\rm Pa}$] & $\{1, 5, 10\}$ \\
Bot. active $p$-level $\tilde{p}_b$\ [$10^5\,{\rm Pa}$] \qquad \qquad & $\{1, 2, 5, 10, 20\}$ \\
Timestep size $\Delta t$\ \ [s]      & \{12, 6, 4.5\}        \\
Viscosity order $\mathfrak{p}$    & \{1, 2, 4, 8, 16\}      \\
$\nu_{16}$ viscosity coeff.\  [$R_p^{16}\,\tau^{-1}$]\       & \{$10^{-48}$, $10^{-51}$, $10^{-53}$\}     \\ 
Robert-Asselin coeff. $\epsilon$  & $ \{0.002, 0.02, 0.2\}$  \\
\hline
\end{tabular}
\caption{Range of numerical parameters for the simulations in this work.}
\label{tab:num_params}
\end{table}

We have also checked the effects of different viscosity order $\mathfrak{p}$ and coefficient $\nu_{2\mathfrak{p}}$ on the flow dynamics.
In general, employing higher values of $\mathfrak{p}$ and correspondingly lower values of $\nu_{2\mathfrak{p}}$ (which fix the energy dissipation rate at the truncation scale) more effectively shields the large-scale flow structures from numerical dissipation and permits the overall kinetic energy spectrum to converge, as was observed in Refs.~\cite{ChoPol96a,SkinCho21}.
In the present work, 16$^{\rm th}$-order hyperviscosity
($\mathfrak{p} = 8$) is chosen because the small-scale structures in deep heating are excessively dissipated for $\mathfrak{p} < 8$ and the kinetic energy spectrum is effectively same up to the dissipation scale for $\mathfrak{p} \ge 8$.
This is consistent with what was found in Refs.~\cite{Choetal03,Choetal08,SkinCho21,SkinCho22}. 

Similarly, the timestep size $\Delta t$ is adjusted when T is varied to ensure numerical stability. 
The $\Delta t$ values in Table~\ref{tab:num_params} (corresponding to \{T341, T682, T1365\} resolutions) are chosen such that the Courant-Friedrichs-Lewy (CFL) number \citep{CFL1928} is below unity, with typical values of $\,\lesssim 0.2$. 
for the entire duration of the simulations.
Finally, we have performed simulations with different values of the Robert--Asselin filter coefficient $\epsilon$, which suppresses the computational mode from the second-order accurate leap-frog scheme employed \cite{Robert66, Asselin72}. 
In this work, $\epsilon = 0.02$ is used because the value prevents the simulation from suffering instability while minimizing artificial damping due to its temporal smoothing.  This value is consistent with that found for {\hd} in Ref.~\cite{ThraCho11}.

\section{D. Additional visualizations}
\begin{figure*}
\centering
\includegraphics[trim={0.0cm 0.0cm 0.0cm 0cm}, clip=true, width=0.67\textwidth]{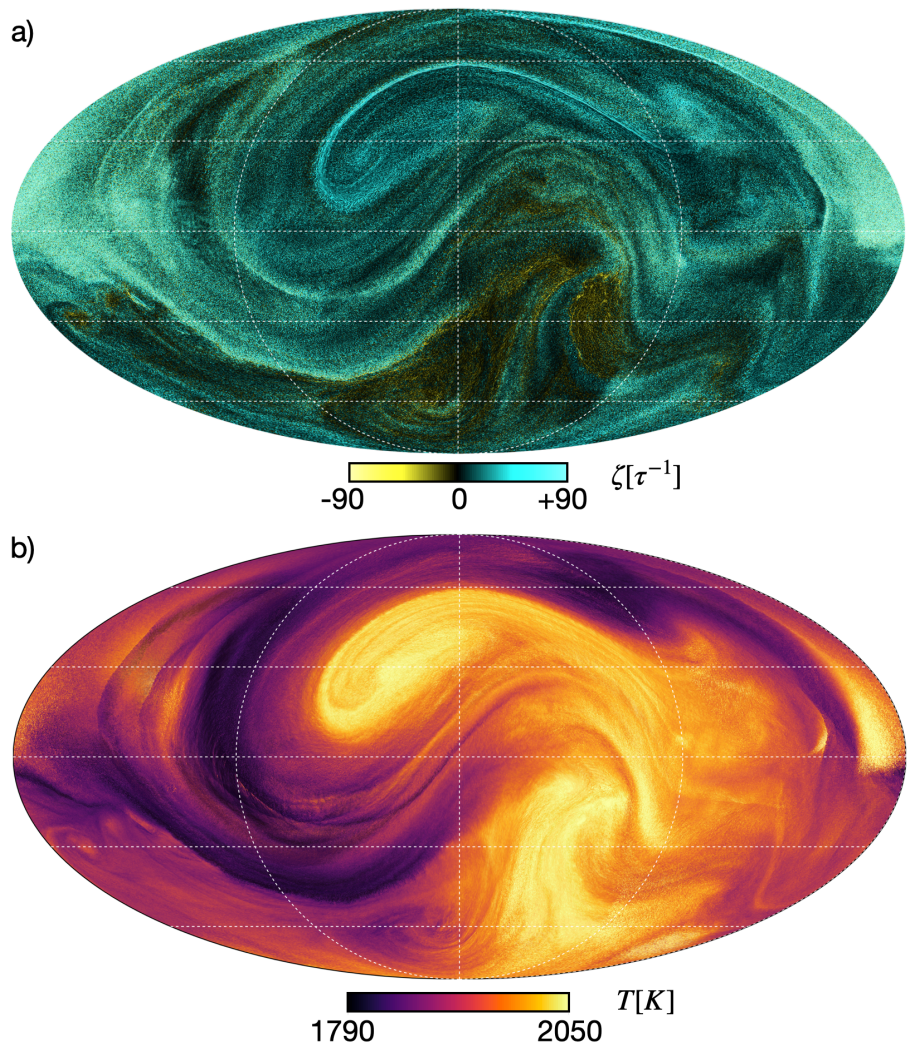}\\
\caption{
High-resolution version of Fig.~1 in the main text.
Vorticity $\zeta$ and temperature $T$ fields (a and b, respectively) of a hot, 1:1 spin--orbit synchronized exoplanet atmosphere with deep heating at time $t = 155$, in units of planetary rotation period~$\tau$.  The fields are in Mollweide projection, centered on the substellar point at the $p = 10^5$\,Pa level; the western and eastern terminators are on the left and right of the dayside, respectively (dashed circle). The generic response to deep heating is shedding of a large vortex, spinning away westward from the substellar point to the north or south, depending on the hemisphere in which the shedding occurs.  The temperature tracks the vorticity very closely.  The shedding is also accompanied by {\it bursts} of small-scale storms and very sharp vorticity and thermal fronts, which are layered and sweeping.  The overall result is a much more complex and widespread mixing of the temperature outside the vortices (as well as the usual transport in their interiors)---seen only sporadically in the atmosphere without deep heating \cite{Choetal21,SkinCho21}.  Here it is the principle state.
\vspace{-0.5cm}
}
\label{fig:fig1_L}
\end{figure*}

\begin{figure*}
    \includegraphics[width = 0.67\textwidth]{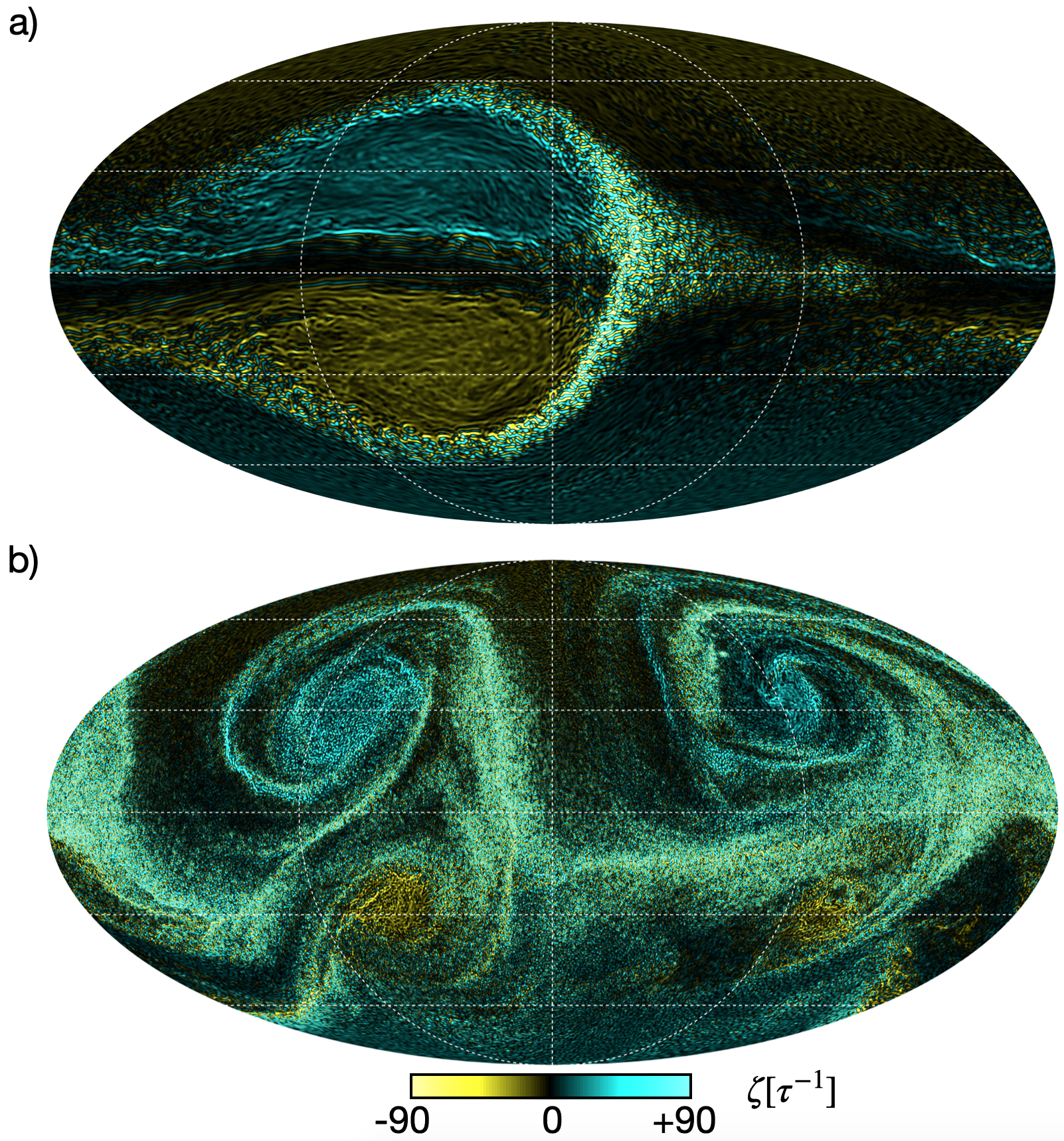}
    \caption{
    High-resolution version of Fig.~3 in the main text.
    $\zeta$ fields at the $10^5$\,Pa level with shallow and deep heating (a and b, respectively), showing the flows dominated by different organized structures (after $t \approx 10$).
    In a) the flow is dominated by a modon and a much weaker antimodon, whereas in b) the flow is dominated by a quartet of cyclones in a von K\'arm\'an vortex street-like formation.  In the latter, the cyclone quartet exhibits a quasi-periodic life cycle (of period $\sim$10) that begins with the flow similar to that seen in  Fig.~\ref{fig:fig1}a. 
    \vspace{-0.5cm}}
    \label{fig:fig3_L}
\end{figure*}

A movie showing the periodic generation of large cyclonic storms from the simulation in Fig.~3b of the main text is available at \cite{youtube_link}. 
The movie shows the vorticity field $\zeta$ ($\equiv {\mathbf{k}} \cdot \nabla\!\times\! \mathbf{v}$, where ${\mathbf{k}}$ is the local vertical direction and $\mathbf{v}$ is the velocity field) in Mollweide map projection for \wasp\ with deep heating.  The map is centered on the substellar point, ($\lambda = 0$ and $\phi = 0$). The field at the $p = 10^5$\,Pa level is shown for $t \in [125.75, 165.00]\,\tau$, during which $\sim$13 distinct formation cycles can be seen. 
Additionally, Fig.~\ref{fig:fig1_L} and \ref{fig:fig3_L} are high-resolution versions of Fig.~\ref{fig:fig1} and Fig.~\ref{fig:fig3} in the main text.

\begin{center}
\line(1,0){70}
\end{center}


%

\end{bibunit}

\end{document}